# Visualization Analysis and Impedance Analysis for the Aging Behavior Assessment of 18650 Cells

Yihan Shi[1,2,5,a], Qingrui Pan[5,a], Jitao Li[1,2,3]*, Xiaoze Shi[6], Youchang Wang[4], Peng Xiao[1]

[1]*School of Science, Southwest Petroleum University, Chengdu 610500, China*
[2]*Chengdu University of Information Technology, Chengdu,610225, China*
[3]*School of Precision Instruments and Optoelectronics Engineering, Tianjin University, Tianjin, 300072, China.*
[4]*School of Aeronautic Science and Engineering, Beihang University, Beijing 100191, China.*
[5]*Hame Technology CO.,LTD, Shenzhen 445414, China.*
[6]*Shenzhen BAK Power Battery CO.,LTD, Shenzhen 445414, China.*
[a]*Authors have equal contribution to this work*
*Corresponding email: jtlee@swpu.edu.cn

**Abstract：**

  This work presents a comprehensive study on the aging behavior of 18650-type lithium-ion batteries, focusing on the uneven intercalation of lithium ions during fast charging processes. It introduces a novel approach using color visual recognition technology to analyze color changes in the graphite anode, indicative of lithiation levels. The study employs X-ray diffraction (XRD) and Distribution of Relaxation Time (DRT) techniques to validate and analyze the observations. The study emphasizes the significance of electrode impedance, the positioning of battery tabs, and electrolyte distribution in influencing the aging dynamics of lithium-ion batteries. Furthermore, the paper presents an innovative impedance Transport-Line Model, specifically developed to capture the evolution of polarization impedance over time. This model offers a deeper understanding of the internal mechanisms driving battery aging, providing valuable insights for the design and optimization of lithium-ion batteries. The research represents a significant contribution to the field, shedding light on the complex aging processes in lithium-ion batteries, particularly under the conditions of fast charging. This could lead to improved battery performance, longevity, and safety, which are critical for the wide range of applications that depend on these energy storage systems.

## 1. Introduction

  Currently, the fast-charging capability of lithium-ion batteries is a critical factor that hinders the complete transition of automobiles to full electrification as excessively high current density can accelerate the aging process of lithium-ion batteries, and in severe cases, even pose safety risks [1-7]. It involves various aging mechanisms, such as lithium plating, electrode material detachment, material lattice collapse, and electrolyte side reactions, all contributing to the loss of active lithium [8-11]. The heterogeneous origins of battery aging may contribute to uneven degrees of aging at different locations throughout the lifespan, which is undoubtedly fatal. It is well known that lithium insertion reactions and other side reactions in lithium-ion batteries are strongly influenced by temperature and the concentration of lithium ions [12]. Differences in the electrolyte content of various electrode standard units, along with temperature gradients caused by variations in structural heat dissipation and impedance [13-17], can give rise to significantly different reaction rates. This discrepancy in reaction rates leads to an uneven degree of aging, ultimately triggering a domino effect of cell failure.

Generally, the relationship between lithiation and de-lithiation of lithium-ion battery anodes during charging and discharging can be indirectly characterized by capacity-voltage data. For instance, Jeff Dahn leveraged the characteristics of Ohmic polarization during constant current charge-discharge processes to separate the polarization voltage caused by internal resistance (CV) from the overall cell voltage, obtaining the shift voltage (SV). The rate of change of SV (SVC) was then employed as an indicator for anode lithium plating [18]. Hisashi utilized the differential voltage (dV/dQ) curve to analyze the degradation of a lithium-ion battery composed of a manganese-based cathode and a graphite anode during discharge [19]. However, the voltage signal is the average potential of the electrodes acquired from the positive and negative terminals at both ends of the cell. In other words, this voltage signal is a coupled result of the electrode itself and the overall polarization state, which makes it challenging to characterize the aging of anode and cathode. Fortunately, one of the macroscopic manifestations during the aging process of the cell is the change in impedance in various parts of the cell with cycles. This allows us to analyze aging behavior from an impedance perspective. Therefore, studying the changing patterns of impedance in different parts of the cell throughout its entire lifecycle can provide a better understanding of the aging characteristics. Electrochemical Impedance Spectroscopy (EIS) has proven to be a powerful tool for characterizing electrochemical systems [20-22]. Among the various EIS analysis methods for lithium-ion batteries, the Distribution of Relaxation Time technique (DRT) [23-26] and appropriately constructed Transmission-Line Models (TLM) [27-30] are considered ideal approaches. The impedance growth behavior of the cell was analyzed by DRT. The results obtained from the DRT are fitted as part of the initial values required for fitting the new transmission line model. Based on this fitting method, the increase behavior of impedance during cell aging is analyzed systematically and accurately.

In addition to electrochemical characterization methods, physical characterization methods are also essential as they provide intuitive insights. However, characterizing lithium, the lightest metal element, poses considerable challenges. Currently, the distribution of lithium embedded in the negative electrodes of cylindrical batteries can only be obtained by high-energy ray detection methods such as neutron diffraction and synchrotron radiation. However, these detection resources are very scarce, which has forced the search for other methods to characterize it. Commercial lithium-ion batteries often use graphite as the negative electrode. During the charging process, lithium ions are intercalated into the graphite layers, forming $Li_xC_6$ compounds. The graphite anode colors at various lithium intercalation depths: grayish-brown (C, $LiC_{72}$), blue ($LiC_{36}$), red ($LiC_{12}$), and golden ($LiC_6$). As lithium intercalation progresses, $Li_xC_6$ undergoes corresponding color changes, which allows for optical analysis of lithium distribution on the electrode. For example, F. Grimsmann and Pfluger et al. [31,32] extensively documented the color changes during graphite lithiation from $LiC_{36}$ to $LiC_6$. Thomas et al. [33] introduced an in-situ optical microscope capable of simultaneously recording images and electrochemical data of the battery cross-section, investigating lithium metal deposition mechanisms induced by edge effects. Balazs [34] estimated the approximate Li content of the "suspended" negative electrode by utilizing the color changes during the three stages of graphite lithiation. Yang et al. [35] achieved in-situ analysis of lithium concentration distribution during graphite lithiation and the resulting uneven stress distribution using an optical imaging system. We creatively proposed the concept of "Color-Voltage" to quantify the Polarization voltage at different locations. Furthermore, Weng et al. [36] introduced a new approach using in-situ X-ray diffraction (XRD) to monitor real-time changes in the peak shape and position of the (002) diffraction peak of graphite during lithiation. Combining the color changes during graphite lithiation with the variations in the (002) diffraction peak of XRD allows for a more quantitative assessment of the lithiation state of graphite.

In this study, DRT analysis on the EIS obtained from 18650 cylindrical cells (cathode: NCM811, negative electrode: Graphite) at different cycling numbers was conducted to investigate the changes in the polarization impedance across various components of the cell as it goes through cycles. as it goes through cycles. To better understand the EIS results, a modified Transport-line impedance model of full cell is proposed. This model is used to fit the Nyquist plot of the cell and extract impedance parameters such as electronic conduction impedance, lithium-ion transport impedance, $PF_6^-$ transport impedance, contact impedance, and diffusion impedance for both the positive and negative electrodes over the lifespan of the cell. Additionally, cells from the same series but with varying cycling numbers are subjected to high-rate charging to intensify uneven lithiation of the electrodes. After charging, the cells are disassembled, and the uneven color distribution of the electrodes, along with XRD diffraction images, is utilized to assess the lithium distribution at different positions. Finally, a cell that has gone through 700 cycles700 cycles is selected for disassembly. The cathode is divided into sections and assembled into coin cells. DRT analysis is then performed on the impedance spectra of these coin cells. Discussions are carried out based on impedance and lithium distribution to assess the localized uneven aging throughout the entire life cycle of the cell.

## 2. Expriment
## 2.1 Electrochemical Impedance Spectroscopy (EIS) and Electrochemical Performance Testing

In this study, a commercial 18650 cell with NCM811 and graphite as the cathode and anode, respectively, was used. The nominal capacity is 2500mAh. Initially, EIS spectra were obtained at different temperatures (-15°C, -5°C, 5°C, 15°C, 25°C) under a 50% State of Charge (SOC) before cycling tests. Subsequently, the cell was aged for 700 cycles with constant current−constant voltage (CC−CV) charging and CC discharging, and EIS testing was conducted under a 50% State of Charge (SOC) every 100 cycles (The test instrument is Zahner IM6e. The test protocol is shown in Fig.S1(a)). The cell undergoing 700 cycles was disassembled following discharging to 0% SOC in the argon-filled glovebox. The designated positions on the electrodes were then cut into small circular electrodes. The cut pole piece is used to assemble the full coin cell. After assembly, the coin cells were left to sit for 12 hours before EIS measurements. The amplitude of the AC was 1 mA and the frequency was varied from 10 kHz to 5 mHz.

To better understand the loss of material in the positive and negative electrodes during the cycling process, a copper-lithium-plated (Cu@Li) reference electrode was used to calibrate the potential of the positive and negative electrodes in the same system and batch of cells during the first charging cycle [37]. Fig.S1(b) illustrated the preparation of the Cu@Li reference electrode.

**2.2 Electrode Morphology Analysis: Optical Analysis and XRD Characterization**

Thiry-five cells from the same batch were charged with 0.25A to a specified voltage ranged from 3.6V to 4.3V in intervals of 0.02V and then kept until the cell current was less than the cut-off value of 5mA. Afterward, the cells were immediately disassembled in the argon-filled glovebox, and the color of the anode at different voltages was recorded to obtain the color-voltage correlation by a color recognition system (KEYENCE, CAH-500MX, as shown in Fig.1(a). The shooting results are seen in Fig. 1(b). The low charging current was designed to mitigate the influence of uneven polarization. Subsequently, cells with different cycle numbers were charged with 20A to 4.2V, and then immediately disassembled in the argon-filled glovebox to obtain the color distribution of the electrodes. Fig.1c shows a schematic photo of the color recognition of part of the Anode after a 20A charge. The color recognition process involves capturing grayscale photos with a black and white CCD camera under 8 different wavelength LED light sources. These photos are then synthesized to obtain color photos, and the grayscale values under different wavelength light sources can quantify color differences (as shown in the Fig.1(a). Finally, 30 samples with the size of 1×2 cm² were cut off from each electrode and then sealed on glass slides using polyimide tape to conduct XRD measurements immediately, as shown in Fig.1(c).

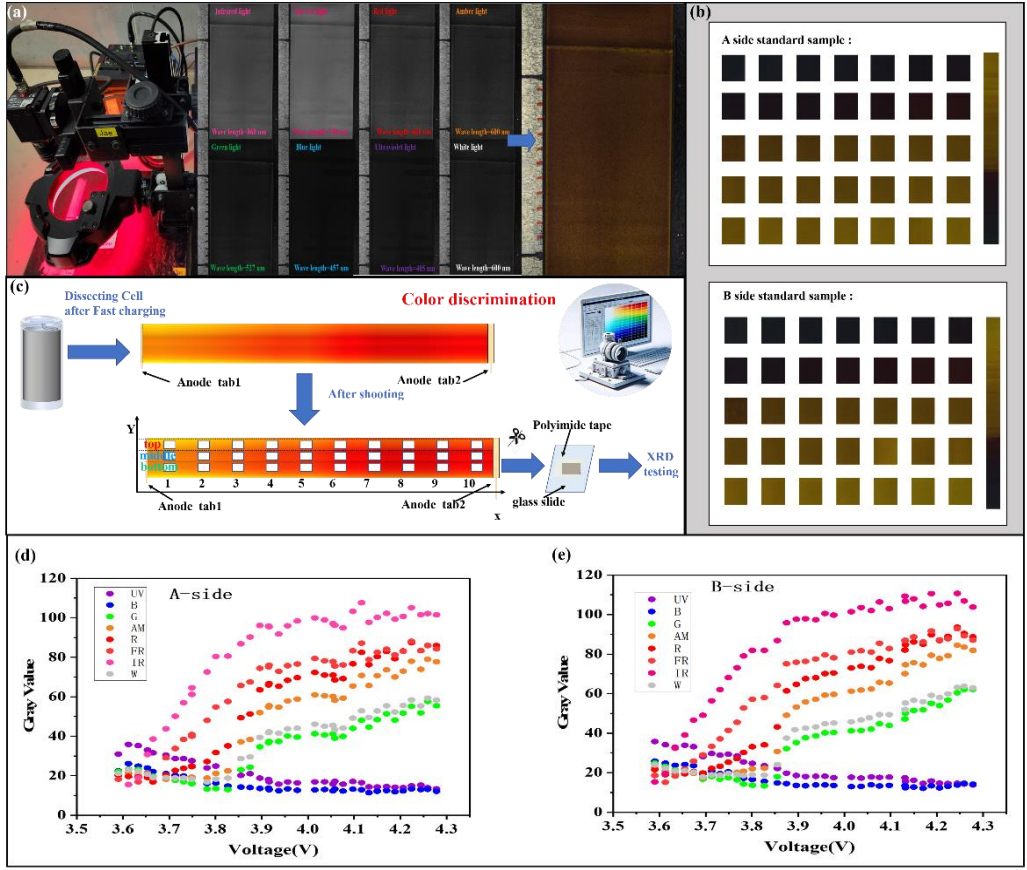

Fig.1. (a) The recognition solution of the color recognition system, (b) color photos of the A-side and B-side of graphite anode under different voltages. (d), (e) Gray values of A and B sides of the graphite anode of 18650 battery under different open circuit voltages under different wavelength light sources.

## 3. Analysis of Impedance Behavior During the Aging Process of 18650 Cells
## 3.1 DRT Analysis During the Aging Process

The analysis of EIS typically involves establishing an Equivalent Circuit Model (ECM) based on the quantity and origin of all processes contributing to the cumulative impedance. However, it becomes challenging to use conventional equivalent circuits along with Nyquist plots to determine the parameters of components within the equivalent circuit due to the significant coupling relationship between them resulting from similar relaxation times of certain physicochemical processes. By contrast, the Distribution of Relaxation Times (DRT) analysis proves to be effective [24, 26].

DRT analysis is based on an equivalent circuit as shown in Fig.2. The circuit comprises an inductive component $L_0$ in the ultra-high-frequency range, an ohmic resistance $R_0$ in the high-frequency range, and in the mid-high-frequency range, a parallel combination of resistance R and capacitor C (R//C element) represents an interfacial polarization process. Multiple polarization processes are represented by the series connection of different R//C elements.

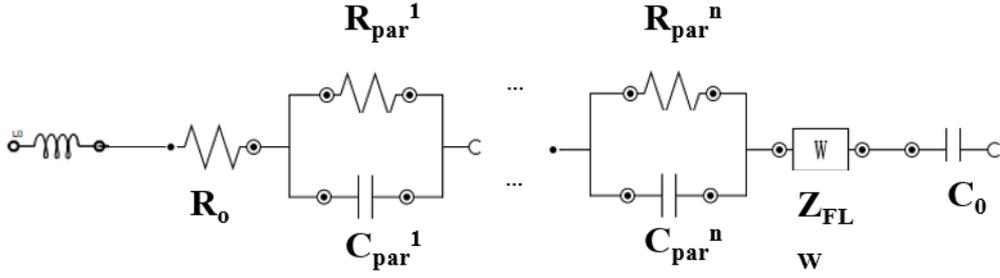

Fig.2. The equivalent circuit diagram for Distributed Relaxation Time (DRT) analysis

Additionally, the concentration polarization overpotential caused by lithium ion diffusion in the low-frequency range lags behind the current. Therefore, the relationship between current and potential resembles a capacitive response, and this process can be represented by a serially connected capacitor $C_0$ in the $Z_{FLW}$ model [36]. The relationship between the total impedance $Z_{total}$ and the polarization processes is given by Equation (1) [36]:

$$Z_{total} = jwL_0 + R_0 + \sum_{i=1}^{n} Z_{RC^i} + Z_{FLW} + \frac{1}{jwC_0} \quad (1)$$

In this context, $L_0$ represents the UHF reactance part of cell, $R_0$ represents the ohmic component in the total impedance, while $Z_{FLW}$ and $\frac{1}{jwC_0}$ are associated with the diffusion effects in the low-frequency range. Additionally, $Z_{RC^i}$ represents the impedance of each RC element and can be expressed by the following equation (2):

$$\sum_{i=1}^{n} Z_{RC^i} = \sum_{i=1}^{n} \frac{R^i}{1 + jwR^iC^i} = \sum_{i=1}^{n} \frac{R^i}{1 + jw\tau^i} \quad (2)$$

Where $R^i$ and $C^i$ represent the resistance and capacitance values, respectively, of a specific RC element, and $\tau^i$ is the time constant of that RC element. The numerical value is the product of the resistance and capacitance [37],

$$\tau^i = R^i \quad (3)$$

When there are a sufficient number of different interface processes in the battery, the discrete distribution in Fig.S2 (c) will become a quasi-continuous distribution of relaxation resistance peaks. Therefore, the time impedance distribution function γ(τ), satisfying the following equation (4), is introduced:

$$Z_{total} = jwL_0 + R_0 + \int_0^\infty \frac{\gamma(\tau)}{1 + jw\tau} d\tau + Z_{FLW} + \frac{1}{jwC_0} \quad (4)$$

If γ(τ) contains multiple peaks of relaxation time distribution, comparing with equations (1) and (2), it is evident that the area of a single peak corresponds to the polarization resistance $R^i$ of the ith polarization process.

According to previous studies [23, 24, 37], the polarization losses in lithium-ion batteries, apart from the ohmic impedance $R_0$ related to ion liquid-phase conduction in the electrolyte and electronic conduction in the electrode, can be roughly divided into five parts: (1) Contact resistance ($R_{Contact}$), (2) Passivation film processes (SEI and CEI, $R_{SEI}$, and $R_{CEI}$), (3) Negative electrode charge transfer resistance ($R_{CT,A}$), (4) Cathode charge transfer resistance ($R_{CT,C}$), and (5) Diffusion resistance ($R_{Diffusion}$). These processes can be arranged in order of time constant τ, from small est to largest: $R_{Contact}$ < $R_{SEI}$ & $R_{CEI}$ < $R_{CT,A}$ < $R_{CT,C}$ < $R_{Diffusion}$ [37-43]. In the Distribution of Relaxation Times (DRT) curve, the peak in the high-frequency range (near kHz) corresponds to the contact resistance ($R_{Contact}$). $R_{Contact}$ is caused by the electrical contact between the active material and the current collector Its magnitude remains relatively constant with temperature but varies with the State of Charge (SOC) due to the expansion stress in the electrode. The distribution of relaxation times in the contact impedance can be explained based on the research by Zhou et al. [24,25], who pointed out that contact impedance is not simply a resistance but a parallel combination of resistance and capacitance, manifested as a high-frequency semicircle in the Nyquist plot which is caused by the variability in the gap between the current collector and the active material. $R_{SEI}$ is formed during the transport of lithium ions through the Solid Electrolyte Interphase (SEI) film. A passivation film (CEI) is also formed on the surface of the ternary lithium material cathode, similar to the SEI on the negative electrode surface [43,44]. The relaxation time constants of CEI and SEI may overlap [45]. Additionally, the activation of $R_{CT}$ occur in the mid-frequency range and is related to the activation of Faradaic reactions at the electrode/electrolyte interface. Therefore, the size of $R_{CT}$ is also temperature-dependent, with $R_{CT,A}$ having a smaller time constant compared to $R_{CT,C}$ [37,45,46]. At room temperature, the time constants of A and B are similar, but they can be distinguished more easily at low temperature. Finally, $R_{Diffusion}$ is associated with the diffusion of lithium ions in the electrolyte and solid active particles [49]. In summary, these polarization resistances exhibit different dependencies on the State of Charge (SOC). Contact and film processes, such as R0 and $R_{SEI}$, are independent of the concentration of lithium ions in solid active particles and therefore do not change significantly with SOC. By contrast, charge transfer and diffusion processes, such as RCT and $R_{Diffusion}$, are highly dependent on SOC [23,24]. On the other hand, the activation energy of polarization resistances is also related to the physical mechanisms. An Arrhenius fit can be performed to analyze the changes in each polarization resistance with temperature, with the slope representing the activation energy magnitude for the polarization process [46,47]. To accurately decompose the relaxation distribution peaks of impedance, this study conducted DRT analysis on the EIS spectra of cells after 240 cycles at 50% SOC and at different temperatures, as shown in Fig.3(a-c). The specific polarization processes represented by each relaxation distribution peak were determined based on activation energy and relaxation time.

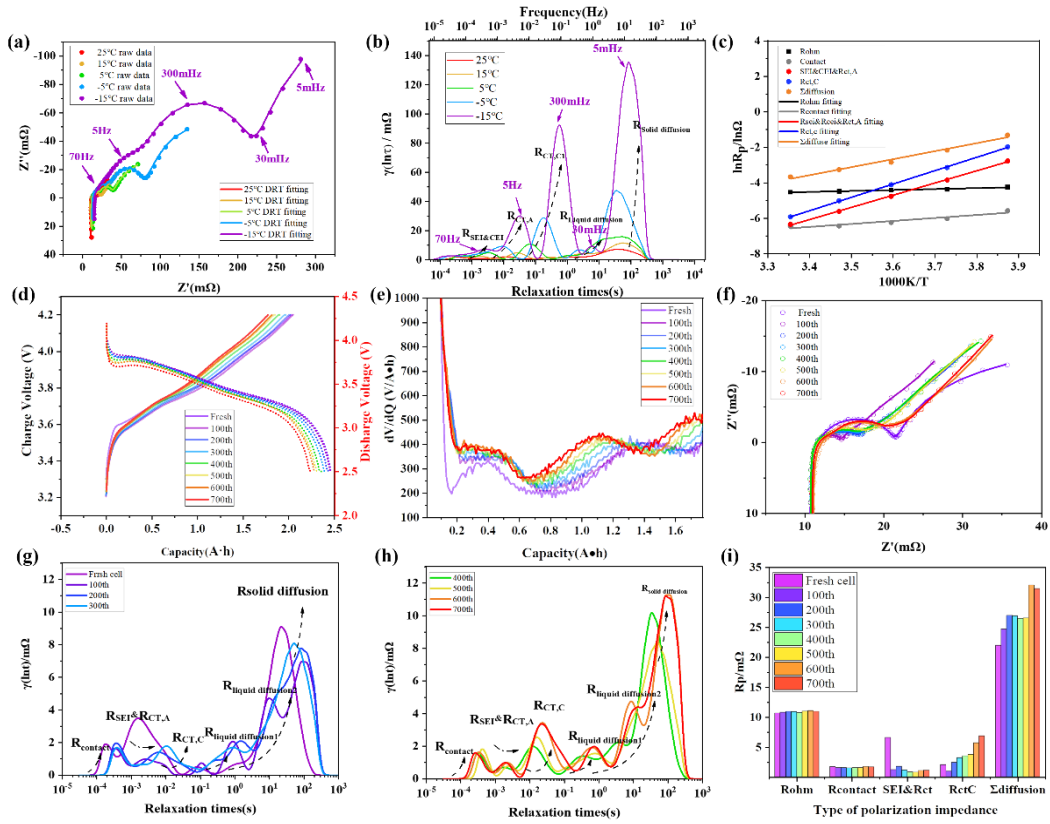

Fig.3. (a) EIS spectra of the cell at 50% SOC after 240 cycles at different temperatures(Raw EIS data (scatter points) and DRT fitting data (lines)), (b) Variation of relaxation distribution peaks with decreasing temperature, (c) Arrhenius fitting of each polarization process at different temperatures,(d)Charge-discharge curves of the cell at various cycling nodes and SEM of anode, (e) dV/dQ curves at each cycling node; (f) Raw EIS data(scatter points) and DRT fitting (lines) at room temperature and 50% SOC after different numbers of cycles, (g) Variation of relaxation distribution peaks during aging(0-300 cycles), (h) Variation of relaxation distribution peaks during aging(400-700 cycles), and (h) Changes in the magnitude of each polarization process during the aging process.

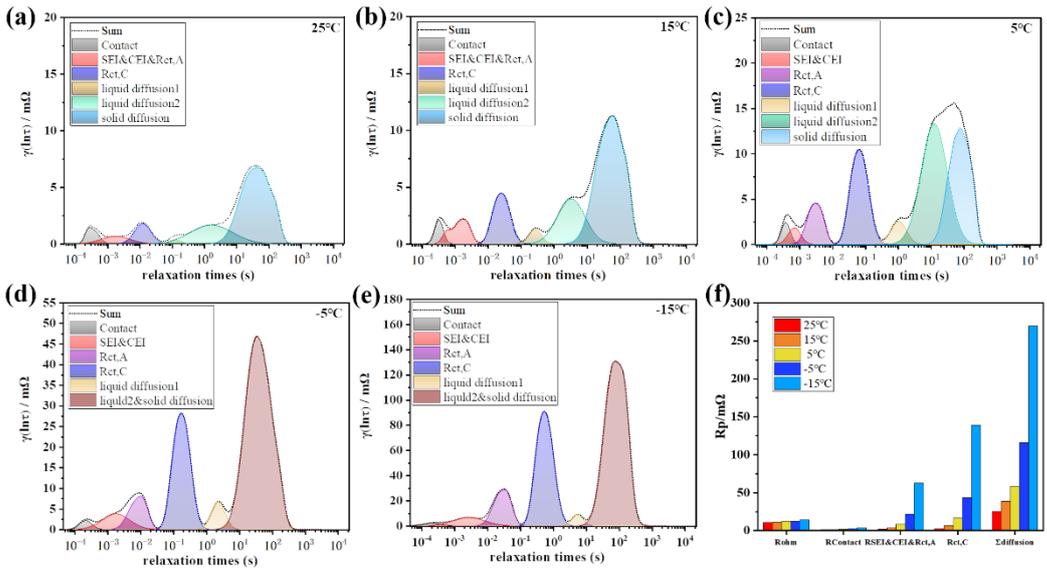

Fig.4. (a-e) Changes in the magnitude and variation of each polarization resistance with decreasing temperature, (f) Peak area of each relaxation peak at different temperatures.

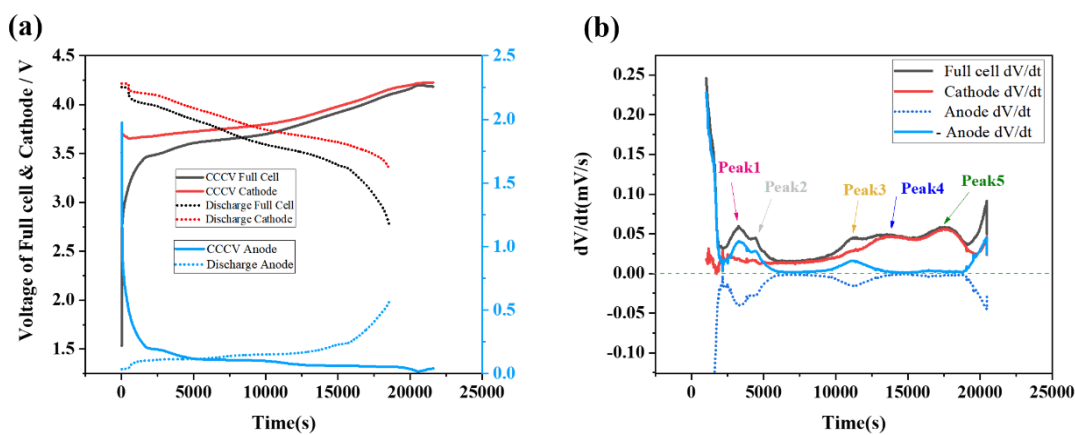

Fig.5. (a) Voltage of Cathode Vs Li/Li+(Cu@Li electrode) and anode Vs Li/Li+ (Cu@Li electrode) during first charging(0.5A), (b)dQ/dV of anode and cathode during first charging.

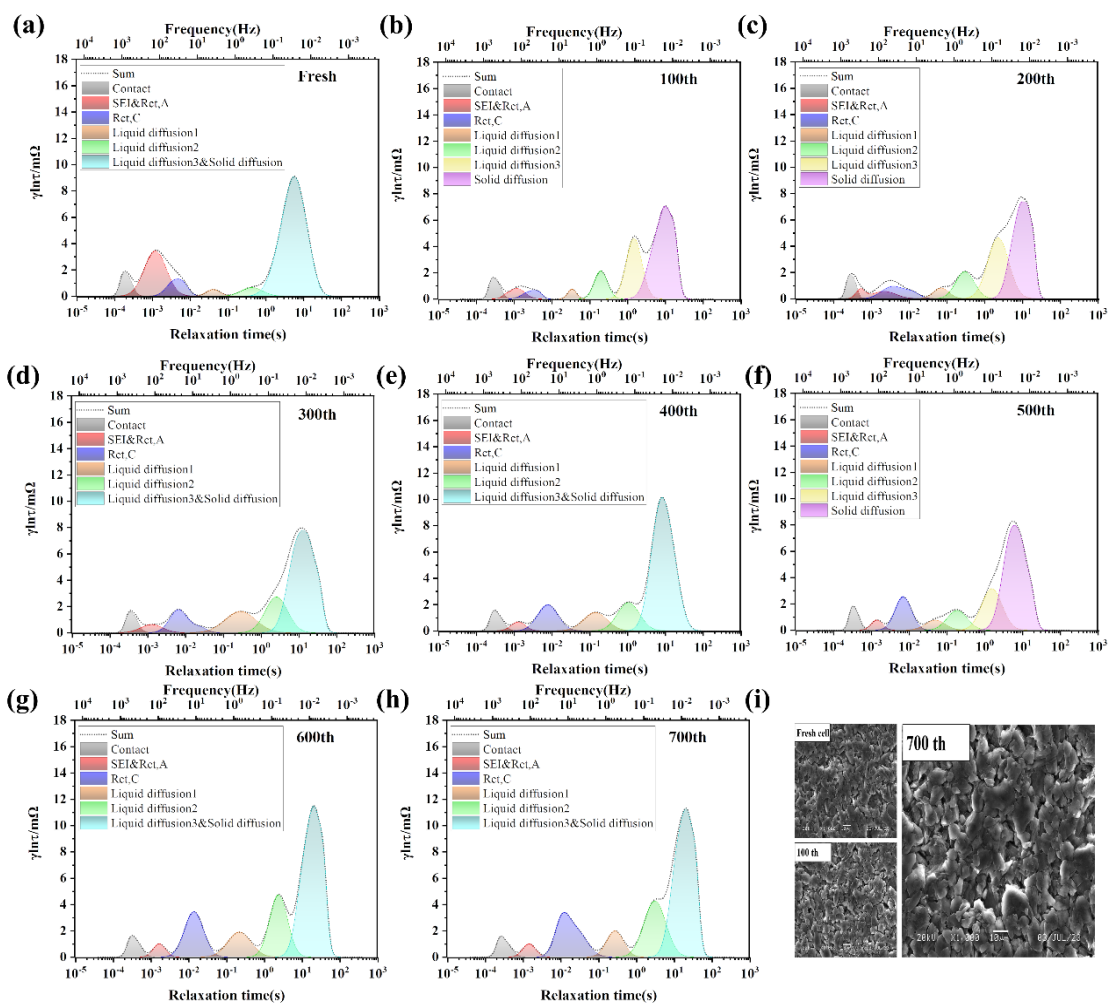

Fig.6. The Distributed Relaxation Time (DRT) distributions of the cells at 50% State of Charge (SOC) and room temperature after different numbers of cycles, (a-h) DRT distribution for the Fresh cell, 100 cycles, 200 cycles, 300 cycles, 400 cycles, 500 cycles, 600 cycles, and 700 cycles (i) DRT distribution after 700 cycles, (i) SEM images of the anode.

It can be observed from Fig.3(a) and Fig.S3(a), that the residuals of the Kramers-Kronig relation for each impedance data are all less than 2%, indicating their linear effectiveness [48] and DRT analysis can be further performed [47, 48]. The fitting errors of the DRT are shown in Fig.S3(b), all of which are less than 0.2%. Fig.3(b) illustrates the variation trends of each polarization process at different temperatures, ranging from low to high frequencies: $R_{Contact}$, $R_{SEI}$ & $R_{CEI}$, $R_{CT,A}$, $R_{CT,C}$, $R_{Diffusion}$., The results of DRT peak division at different temperatures are shown in Fig.4. Firstly, it should be noted that due to the non-uniform internal state of the battery, such as different electrolyte contents on the inner and outer sides of the winding core, different pressures, and even variations in particle sizes of cathode and anode materials, there may be significant differences in the relaxation times of certain interface polarization processes within various polarization processes. In other words, a certain process may exhibit more than one relaxation peak, a phenomenon also observed in previous studies [45-47], though not explicitly elucidated. The liquid diffusion process, as illustrated in Fig.4(d), exhibits two relaxation peaks, herein defined as $R_{liquid\ diffusion1}$ for the shorter time constant and $R_{liquid\ diffusion2}$ for the longer time constant. This distinction arises from the differential wettability of the electrolyte between the inner and outer layers of the cylindrical electrode core, a point to be further discussed in subsequent sections. This phenomenon becomes more pronounced at lower temperatures due to the increased viscosity of the electrolyte, accentuating the differences in wettability and leading to greater variations in impedance growth during the liquid diffusion process. At slightly elevated temperatures, the relaxation time for the anodic charge transfer process ($R_{ct,A}$) and the relaxation processes associated with the Passive film (SEI&CEI) are closely aligned. However, at lower temperatures, $R_{ct,A}$ and the relaxation processes of the Passive film are separated, as depicted in Fig.4(a-e). As the temperature decreases, the relaxation times of various processes increase. With decreasing temperature, the relaxation peaks of these processes may exhibit characteristics of separation or re-coupling in terms of relaxation time. This provides guidance for us to perform peak identification. Fig.4(f) depicts the change in peak area with temperature, and from here, we observe typical exponential growth, which is in full accordance with the Arrhenius relationship, confirming the correctness of peak segmentation. Given the coupling of ($R_{ct,A}$) and the Passive film process at room temperature and the coupling of Rliquid diffusion and Rsolid diffusion at low temperatures, for the sake of minimizing unnecessary errors, the relaxation processes were not separated when calculating impedance values through integration. Instead, ($R_{ct,A}$) and the Passive film process were treated as an integrated charge transfer process (SEI & CEI & $R_{ct,A}$), and liquid diffusion2 and solid diffusion were amalgamated into a comprehensive diffusion process (Σdiffusion)

Considering the changes in various relaxation peaks, it is evident that, apart from the $R_{Contact}$ process, all other processes exhibit a strong temperature dependence. As the temperature decreases, each polarization strengthens, and the relaxation time also increases. The process most affected by temperature is evident from the comprehensive analysis of the variations in each relaxation peak that, excluding the $R_{Contact}$ process, all other processes exhibit strong temperature dependencies. As the temperature decreases, polarization processes intensify, and relaxation times increase. It can be observed that the charge transfer process is most affected by temperature, followed by the $R_{Diffusion}$ process. Additionally, due to the increased relaxation time of liquid diffusion at lower temperatures and the faster growth of $R_{liquid\ diffusion2}$ compared to $R_{liquid\ diffusion1}$, at even lower temperatures (-5°C and -15°C), $R_{liquid\ diffusion2}$ and $R_{solid\ Diffusion}$ are coupled together. This coupling is attributed to $R_{liquid\ diffusion1}$ corresponding to the slightly outer side of the core with better electrolyte wettability, while $R_{liquid\ diffusion2}$ corresponds to the slightly inner side with poorer electrolyte wettability. The decrease in temperature has a more significant impact on $R_{liquid\ diffusion2}$, reflecting greater changes. Fig.3(c) presents Arrhenius linear fits based on the natural logarithm values (LnR) of polarization impedances at various temperatures plotted against 1/T. This indicates that the polarization resistances exhibit exponential changes with temperature, conforming to the Arrhenius equation, validating the decomposition of relaxation peaks. As mentioned earlier, due to the coupling of $R_{ct,A}$ with the Passive film process at room temperature and the coupling of $R_{liquid\ diffusion2}$ with $R_{solid\ Diffusion}$ at low temperatures, it is challenging to separate them. Therefore, this study did not perform separate Arrhenius fits for $R_{ct,A}$ with the Passive film process and liquid diffusion with solid diffusion. Instead, the sum of $R_{ct,A}$ and the Passive film process, as well as the sum of liquid diffusion and solid diffusion, were subjected to Arrhenius fitting (red and orange lines in Fig. 3(c), and the results also satisfactorily adhere to the Arrhenius relationship.

Fig.3(d) shows the charge and discharge curves of the battery at each cycle node. From the CC charging curve, it can be observed that polarization initially decreases and then increases. Fig.3(e) is the dQ/dV curve calculated using the CC segment during the charging process in the cycling. Comparing the dV/dt curves of the cathode and anode with the full cell (as shown in Fig.5), it is evident that Peak1 and Peak2 near an Anode potential of 0.2V correspond to the lithium insertion plateaus of graphite in two Potential platform. Peak3, dominated by the negative electrode phase transition (Anode voltage at 0.1V, Cathode voltage near 3.7V corresponding to the NCM's H1-M phase transition), while Peak4 (Cathode voltage at 4.0V) corresponds to the NCM's M-H2 phase transition, and Peak5, where the Cathode voltage is 4.19V, corresponds to the NCM's H2-H3 phase transition [49-51]. It can be observed from Fig.3(e) that Peak4 and Peak5 gradually sharpen, representing lattice losses caused by the oxygen release during the phase transition of the cathode material [52]. The H2-H3 phase transition causes the release of active oxygen in the material, leading to a drastic change in the crystal cell volume, resulting in stress accumulation within the material particles and ultimately causing particle cracking. Secondly, the peak area of dQ/dV represents the capacity of the plateau reaction. It can be seen that both Peak1 and Peak2 undergo significant decay, which is related to the repeated growth of the SEI film [51]. Fig.3(f-h) presents the EIS and DRT fitting of the cells undergoing different cycles as observed in Fig.3(g), (h) The distribution of impedance relaxation peaks changes significantly at different cycling stages. Integrating the areas of the relaxation peaks in each DRT provides the polarization resistance values shown in Fig.3(i), with the decomposition process detailed in Fig.6(a-h). First, except for the $R_{Contact}$ process, which undergoes minor changes, all other processes exhibit significant variations. The black arrows in

Fig.3(g) and (h) illustrate the changing trends of each relaxation peak. Although the change in contact polarization is relatively small compared to other processes, a decreasing trend in its peak area is observed initially, likely due to the increased pressure caused by the expansion of the electrode, resulting in a tighter electrical connection between the current collector and the active material. The later decrease is attributed to the fatigue of the binder due to prolonged physical expansion, leading to an increased distance between the active material and the current collector. Second, in the fresh state, the relaxation peak of $R_{SEI}$ & $R_{ct,A}$ exhibits a wider distribution and larger area. However, after 100 cycles, the relaxation peak of $R_{SEI}$ & $R_{ct,A}$ collapses, and with the continuous cycling, $R_{SEI}$ & $R_{ct,A}$ undergoes a reciprocating decrease and increase. This corresponds to the dissolution of the SEI membrane in the organic solvent during the cycling process, the rupture of the SEI membrane caused by shear forces resulting from internal pressure and expansion stress, an increase in impedance of liquid diffusion leading to concentration polarization, and the synergistic effects of the continuous secondary reactions of electrolyte components exposed on the Anode surface after membrane cracking [38]. Additionally, it can be observed that the height of the relaxation distribution peak of $R_{SEI}$ & $R_{ct,A}$ decreases for most of the time and shifts towards larger time constants. For a single interface within the DRT distribution peak of $R_{SEI}$ & $R_{ct,A}$, it is only necessary to draw a line perpendicular to the relaxation time axis within the relaxation peak. The value of $\gamma(\ln\tau)$ at that point is proportional to the membrane resistance of that interface, where the time constant is the product of membrane resistance and membrane capacitance. Here, the time constant is the product of the membrane resistance and the membrane capacitance. The decrease in height suggests a reduction in the membrane resistance, but the increase in the time constant indicates an increase in the membrane capacitance. According to Zhang's study [20], an increase in the electrochemically active surface area significantly increases the capacitance value. Therefore, the increase in membrane capacitance should be attributed to the growth of the membrane area during the aging process. These signs indicate that the growth process of the SEI film in this cell results in an increased area, but a decrease in membrane resistance. Therefore, the change in the resistance of the SEI comes from the adjustment process of the membrane thickness and area during the cycling process [2].

$R_{ct,C}$ increases due to lattice losses resulting from the oxygen dissolution during the H2-H3 phase transition in the cathode. The initial weakening of $R_{ct,C}$ in the early cycles could be attributed to the material infiltration process and a reduction in the polarization of the RSEI & RCEI processes in the early stages of cycling, as the CEI is also included in the relaxation peak of the RSEI & RCEI processes [49]. Finally, it is worth noting that liquid diffusion involves three relaxation peaks (Fig.6) due to differences in the electrolyte wettability between the inner and outer layers of the electrode [23,24,49]. As the electrolyte is consumed during the cycling process, the polarization impedance of liquid diffusion increases, and the process with a longer relaxation time, liquid diffusion3, couples with the solid diffusion process. For instance, in the fresh state of the battery, the liquid diffusion3 peak and solid diffusion peak are coupled together, indicating that their relaxation times are close. After 100 cycles, they separate, and with further cycling, the liquid diffusion peak gradually couples with the solid diffusion peak. This phenomenon is evident in Fig.3(g) and Fig.6. This process may be owing to the fact that as the cycling advances, the solid-phase diffusion polarization initially increases due to the loss of material structure, and the two are separated by the increased relaxation time of solid-phase diffusion polarization. As the electrolyte is consumed, the liquid-phase diffusion polarization rises, and the two peaks recouple. Finally, as the electrode material structure ages, the solid-phase diffusion relaxation peak deviates from the liquid-phase diffusion relaxation peak once more.

The impedance changes during the cycling process of the battery are closely related to the micro-surface state of the electrodes. Therefore, we selected batteries of the same model and production batch for cycling tests. At different cycling stages, we disassembled the batteries and observed the surface morphology of the negative electrodes at specified locations (as shown in Fig.S4(a)). The SEM results of the negative electrode sheets are shown in Fig.6(i). Firstly, through SEM, it is evident that there is a significant amount of SEI film (darker regions) on the surface of the negative electrode in both fresh and late cycling stages, corresponding to the cyclical growth of the membrane impedance observed in the impedance analysis above. Finally, we observed an interesting phenomenon: on the inner side of the winding core of the electrode sheet (position 3), the distribution of SEI on the surface showed a decrease followed by an increase. This is consistent with the findings of Rong's study [52], which pointed out that the inner circle of the winding core of the battery experiences higher pressure. The expansion stress and pressure generated by the lithium expansion of the negative electrode create shear forces that lead to the rupture of the SEI film at position 3. After the rupture, it continuously undergoes side reactions with the electrolyte, resulting in SEI regeneration. In fact, this is also a major reason for the drying up of the inner circle of the winding core in the later stages of the battery's lifecycle. Additionally, the gases generated by the side reactions between the negative electrode active material and the electrolyte exacerbate the accumulation of internal pressure. Taking the side reaction with DMC and the cathode as an example [56], the reaction equations are given by equations (7) and (8):

$$CH_3OCO_2CH_3 + 2e^- + 2Li^+ \rightarrow 2CH_3OLi \downarrow + CO \uparrow \qquad (7)$$

$$CH_3OCO_2CH_3 + 2e^- + 2Li^+ + H_2 \rightarrow Li_2CO_3 + 2CH_4 \uparrow \qquad (8)$$

**3.3 TLM analysis of impedance behavior during the aging process**

Although DRT analysis provides significant guidance for interpreting EIS, to better analyze the reasons for aging failure using impedance spectra, it is also crucial to use a suitable circuit model to obtain the corresponding impedance parameters. Since the surface of commercial electrode materials is porous, and the current distribution on the electrode is uneven when the three-dimensional electrode is shaped, an equivalent circuit based on a planar electrode structure model is not an ideal choice. According to de Levie's research [29,30], TLM can accurately fit the response in the high-frequency and medium-frequency regions (mainly the transport of ions in the liquid phase and the transport of electrons in the porous electrode conductor, contact impedance, charge transfer impedance, etc.). However, the model proposed in 1960 has an important conceptual flaw, where the electrolyte phase is described by a single resistor element, which fails to produce any relevant liquid diffusion effects. As a result of ion diffusion in the electrolyte phase enclosed in the porous electrode, this model is unable to reproduce the impedance arcs at low frequencies observed in EIS spectra. Therefore, if the diffusion process of ions in the porous electrode significantly affects the impedance of the electrode in the studied system, it is necessary to upgrade the TLM model to simulate this process. In fact, Jože proposed that the transport number of $PF_6^-$ in the process of assisting lithium ion transport is often greater than that of lithium ions, so introducing an anion transport channel into the TLM can effectively address this issue, as shown in Fig.7(a) [27-28]. Jože established a 3D equivalent circuit based on this transport model, as shown in Fig. 7(b), and equivalently transformed it into a planar equivalent circuit shown in Fig.S5. It should be emphasized that although the two circuits do not have a completely equivalent relationship, the impedance output differences between the two circuits are less than one percent when the parameters are within an appropriate range. Definitions of parameters in Fig.7(a) are provided in Table S3 and Fig.S5, and mesh current analysis of the equivalent circuit in Fig.7(c) yields a matrix equation for mesh currents and an analytical form of total impedance, creating feasibility conditions for using this model to fit measured EIS data. For a description of the transport model and ECM, as well as the analytical form of total impedance ($Z_{total}$) mesh currents, refer to the supporting material.

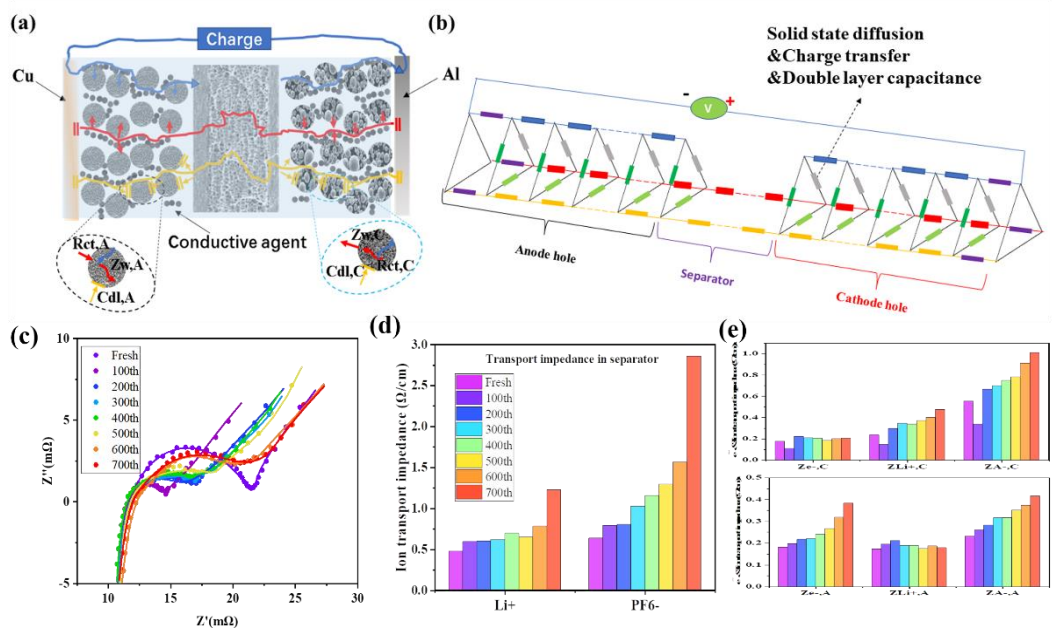

Fig.7. Full-cell transmission line circuit: (a) Schematic diagram of electron, lithium ion, and $PF_6^-$ transmission lines, (b) 3D equivalent circuit of the full-cell transmission line, (c) Planar equivalent circuit of the full-cell transmission line, (c) Nyquist plots of the cell at different cycle numbers (scatter points), at 25°C and 50% SOC, along with their fitting results(lines), (d)and(e) Impedance results of various components fitted using the full-cell transmission line model.

By utilizing the aforementioned full-cell transmission line model to fit the EIS curves of cells at different cycles, in order to reduce the computational complexity of the fitting, the Rct results obtained from the DRT were used to estimate the initial value range of charge transfer impedance in the TLM model. Fig.7(d) shows the TLM fitting results: Firstly, regarding the diffusion impedance and contact impedance, the TLM analysis and DRT analysis yielded consistent conclusions. That is, with the progress of cycles, the negative electrode contact impedance and the liquid-phase resistance within the negative electrode domain separator, as well as the cathode contact impedance and the liquid-phase resistance within the cathode domain separator, along with the diffusion impedance of both positive and negative electrodes, gradually increased. This is attributed to the damage to the structure of positive and negative electrode materials and the drying of the electrolyte due to side reactions [59]. In addition, except for the cell at 100 cycles, the electron impedance of the positive and negative electrodes, $Li^+$, and the non-active anion transport impedance have all increased. Among them, the increase in the electron transfer resistance of the cathode is the most rapid. The reason for the decrease in battery impedance when the cell is cycled for 100 cycles may originate from the activation process of the positive electrode material.

**4 Analysis of Lithium Insertion Distribution in Electrode Sheets**

F. Grimsmann revealed a discrepancy in color between the graphite negative electrode during both charging and discharging, even when they are in the same state of charge, and this color hysteresis phenomenon persists after any waiting time [31]. Therefore, the dependence of the negative electrode sheet color on the lithiation degree of graphite allows us to measure the uneven lithium insertion distribution in the electrode sheets using the color of the disassembled electrode sheets. Additionally, as mentioned earlier, XRD diffraction results of different positions in the electrode sheet can also assist in judging the strength of lithiation in areas with similar colors [36,57]. Of course, due to the non-in situ observation of color in the experiment and the destructive disassembly of the electrode sheets for XRD testing, different battery cells at various cycling stages are needed. This undoubtedly poses a challenge to the consistency of the battery cells. Fig.8(a) displays six cells from the same batch that exhibit similar cycling performance. The maximum difference in cycling retention rate among these six batteries at the same cycling stage is 1.28%, indicating good consistency among these batteries. This batch of batteries can be used to summarize the changes in the uneven lithium insertion distribution during the cycling process of this type of battery cell at different cycling stages. Fig.1(b) show the average grayscale concentration identified from the photos taken on the A-side and B-side at different voltages under different background light sources. The identification points are rectangles with an area of 8x8 pixels. Based on this, linear correlation calculations were performed on the voltage-grayscale of the A-side and B-side, respectively, as shown in Table S2 and Table S3. The three segments with the strongest linear correlation were selected for linear fitting (3.85≥U≥3.6V, IR ray, 4.0>U>3.85V, White ray, 4.3≥U≥4.0V Green ray). As demonstrated in Fig.8(b) and Fig. S7(b), the grayscale values, Color-Voltage, were used to estimate the voltage using the fitted lines.

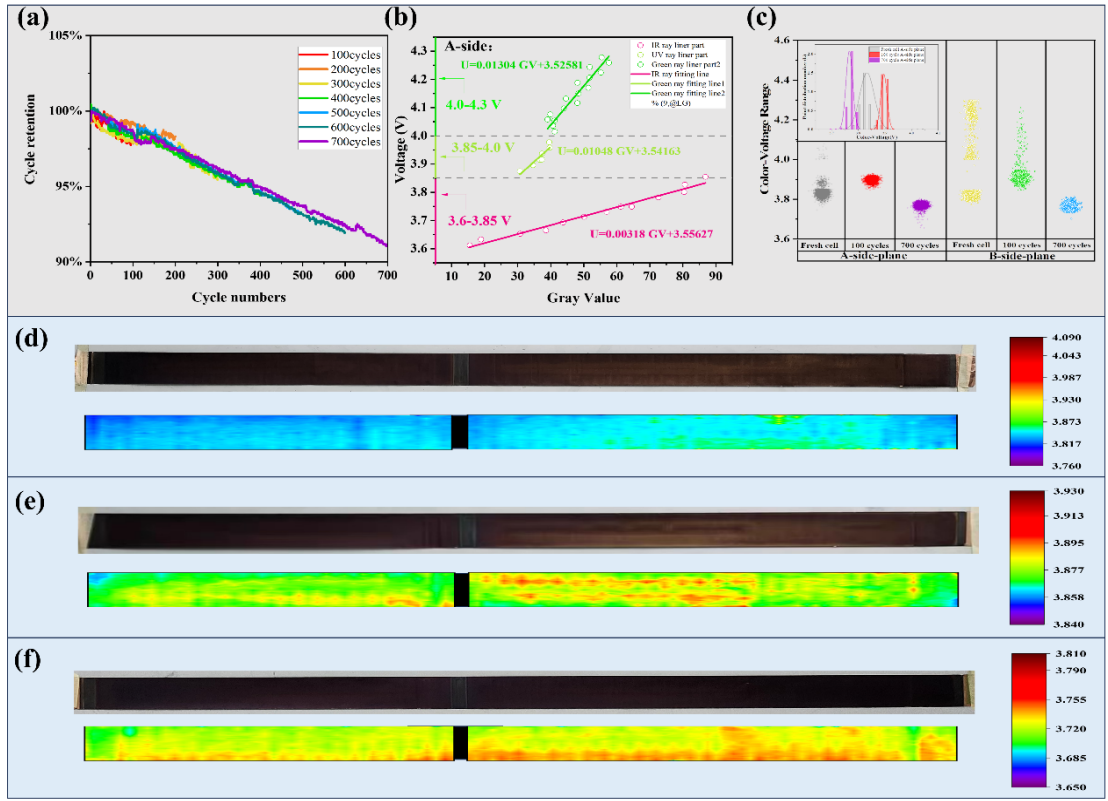

Fig.8. (a) Cycling performance of six cells from the same batch, (b) Fitted curve of the Voltage-grayscale value relationship for the A side of cells,(c) Scatter distribution of Color-voltage,(d-f) Color-Voltage Distribution after Fast Charging for Fresh Cells, 100 Cycles, and 700 Cycles.

Fig.8(b-f) shows the Color-Voltage distribution on the electrode surface (Fresh cell, 100 cycles, 700 cycles), which was obtained by calculating the voltage at various points using the fitted lines in Fig.8(b). The XRD analysis of anode as shown in Fig.S8(a). It should be noted that due to the high charging rate, the electrode slices are not a homogeneous phase within the XRD test range, resulting in multiple diffraction peaks, such as the four states of $Li_xC_6$ diffraction in Fig.S8 (Stage1, Stage2, Stage3, Stage4), where the lithium insertion degree gradually decreases from Stage1 to Stage4, with Stage1 being the $LiC_6$ diffraction peak [19,36].

Firstly, even though the cell voltage reaches 4.2V at the end of charging, the Color-Voltage distribution reveals that, due to the exacerbation of polarization voltage at high rates [18], the majority of color voltages at most positions are in the range of 3.8-3.9V, with only a few positions approaching 4.2V. Of course, with the progression of cycles, the maximum values of Color-Voltage at various locations on the electrode decrease. However, the area of the high potential regions on the electrode increases before decreasing. Consistent with the previously observed trend of total polarization impedance falling before increasing, this results in an overall average increase and subsequent decrease in the degree of lithium insertion in the negative electrode as a function of cycling According to the previous impedance analysis results, the initial decrease in polarization is primarily caused by the SEI film rupturing., while the later increase in polarization voltage is attributed to the increase in electronic conduction resistance, ion transport resistance, diffusion resistance, and the loss of active lithium during the cycling process. Fig.8(c) shows the scatter frequency distribution of Color-Voltage, which further indicates that the distribution range of Color-Voltage gets smaller as the cycle continues. Therefore, it appears that polarization growth is most pronounced in the early phases of the process in the high voltage regions. This is due to the fact that depth of charge and depth of discharge are greater in high voltage zones because of the higher current density in these locations during the cycle charge-discharge process. This causes the areas with high voltage to age more rapidly than those with low voltage. It may be inferred that the B-side of the electrode slice has a higher lithium insertion degree from the color-potential distribution compared to the A-side (as shown in Fig 8(c) and Fig S7). This is related to the formation process of the current collector. The contact surface between the current collector on the B-side and the active material is a rough surface, while on the A-side, it is a smooth surface. Therefore, the contact on the B-side is better. Additionally, the local N/P ratio (the ratio of the contribution capacity of the cathode to the negative electrode within a local region) on the B-side is smaller compared to the A-side, as shown in Fig.9. As a result, the amount of lithium inserted on the B-side is likewise increased.

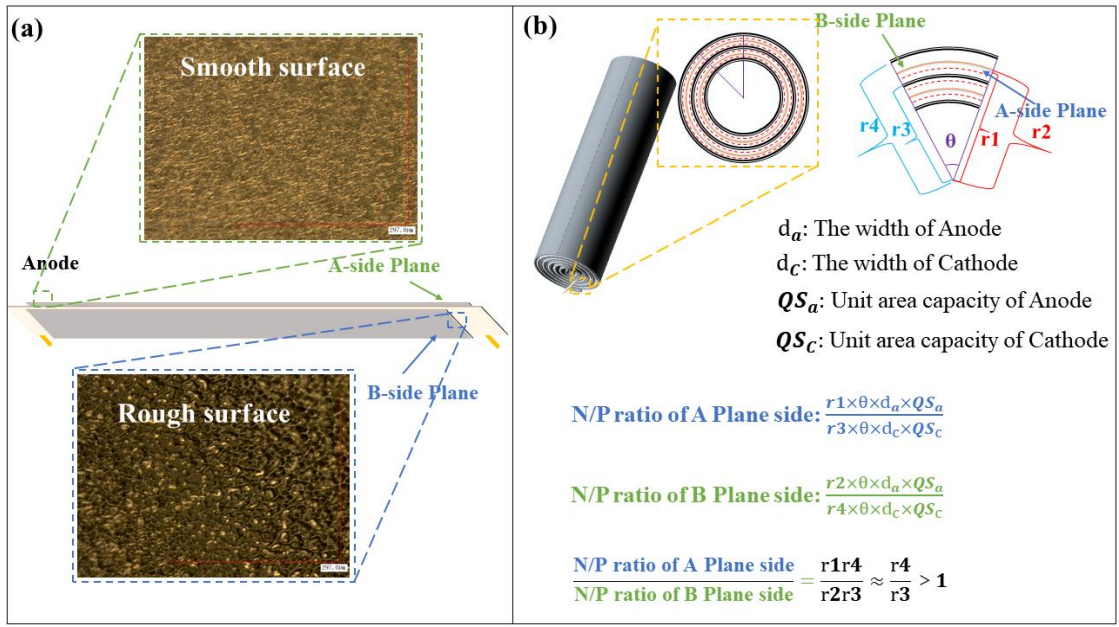

Fig.9. The reason for the higher Color-Voltage on the B-side Plane compared to the A-side Plane. (a) depicts the contact surface between the active material and the current collector copper foil on both the A-side Plane and B-side Plane. (b) provides a comparison of the N/P ratio between the A-side Plane and B-side Plane.

The XRD test results match with the results obtained from Color-Voltage, indicating that the color-potential method can be used to quantify the distribution state of lithiation on the surface of the graphite negative electrode. The XRD test results show that there is no diffraction peak of Stage1 (LiC$_6$) after the later cycles [19]. This is undoubtedly due to the increase in polarization voltage during the cycling process. Furthermore, it is noteworthy that in the X-direction of the electrode slice, there are

different Color-Voltage distribution patterns for cells with different aging degrees:

Following charged at a rate of 20A, the Color-Voltage distribution on the electrode slices reveals a progressive reduction from the outer to inner layers of the jelly roll. The degree of lithium implantation gradually decreases from the outer rings to the inner ones, but it is significantly larger at the extremities of the jelly roll than in the middle，in consistent with D. Petz's findings from neutron diffraction [58,59]. Nevertheless, the impact of the tab on the distribution of lithium insertion is minimal because the cells utilized in this investigation have a double-tab configuration for the anode. In the vicinity of the corresponding position of the cathode tab, there is a slight increase in Color-Voltage compared to the surrounding small area, but it is not significant. The lithium insertion model (Fig.10) is established to provide an explanation. the distribution of the electrolyte is mostly responsible for the color-potential distribution, which has bigger values at the ends than in the middle of the width direction. The steel shell of the cell strongly binds gas produced at both ends of the cell, forcing the gas to concentrate in the middle of the cell. As a result, the electrolyte is pushed toward both ends, leading to a greater degree of lithium insertion at the width ends. This phenomenon intensifies with an increase in gas production during cycling. Secondly, in the length direction, the degree of lithium insertion on the outer side of the jelly roll is greater than on the inner side. This is also attributed to the distribution of the electrolyte. Due to higher pressure on the inner side of the jelly roll, there is less electrolyte, as observed in disassembled cells where the inner side appears drier while the outer side is more humid. Additionally, during the charging process, pressure and expansion of material particles generate shear forces on the negative electrode surface, causing the SEI film to rupture on the material particle surface. The reaction of the electrolyte with the new surface further consumes electrolyte, contributing to the reason for the lack of electrolyte on the inner side [38].

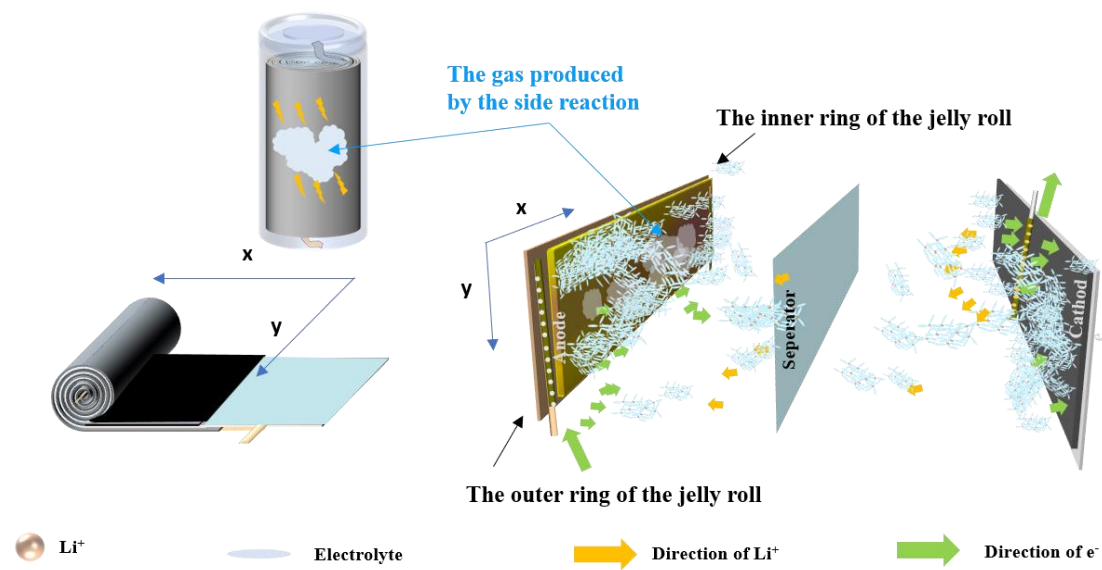

Fig.10. Lithium Insertion Distribution Model for Graphite Anode

## 4. Conclusion

This paper analyzes the variations in the impedance during the aging process of commercial 18650-type lithium-ion cells and the uneven lithium intercalation of lithium ions into the graphite negative electrode. Combining scanning electron microscopy (SEM), charging curve analysis, and impedance analysis, it is concluded that the aging of the cell is primarily influenced by the impedance of the electrode film, diffusion impedance, and the increase in transfer resistance within the cathode, anode and separator. Additionally, the aging process is substantially influenced by the increase in bulk resistance resulting from material loss. The color variation of $Li_xC_6$ in graphite during lithiation is utilized, and based on a color recognition visual system, the distribution of Color-Voltage is obtained. Based on this, a model is presented to summarize the changes in lithium insertion polarization distribution during the aging process of 18650 cylindrical cells. The results are further validated using information from X-ray diffraction (XRD) peak variations. The distribution of Color-Voltage indicates that, under the structure of double negative tabs, the non-uniform lithium insertion behavior in the graphite negative electrode of cylindrical lithium-ion batteries during fast charging is mainly influenced by the wetting of the electrolyte and the synergistic effects of non-uniform pressure inside the jelly roll. Unlike the neutron diffraction method used to obtain the lithium distribution inside the battery cell, the original color-potential method introduced in this paper provides a fast and economical way to evaluate the distribution of polarization voltage inside the battery cell.

**Acknowledgements**
Thanks to Shenzhen Anche Technologies Co., Ltd. for their help in this article in the form of fitting program optimization. Appreciation is extended to Professor Zhou Xing of the National University of Defense Technology for his invaluable expertise and guidance in the impedance spectroscopy analysis aspect of this manuscript.

**Conflicts of Interest**
The authors declare no conflicts of interest.

**Data Availability Statement**

The data that support the findings of this study are available on request from the corresponding author.

**REFERENCE**

[1] W. Cai, Y.X. Yao, G.L. Zhu, C. Yan, L.L. Jiang, C. He, J.Q. Huang, Q. Zhang, Chem Soc Rev, 49 (2020) 3806-3833.
[2] B. Scrosati, J. Garche, Journal of Power Sources, 195 (2010) 2419-2430.
[3] X.B. Cheng, R. Zhang, C.Z. Zhao, Q. Zhang, Chem Rev, 117 (2017) 10403-10473.
[4] M.M. Mench, C.Y. Wang, M. Ishikawa, Journal of The Electrochemical Society, 150 (2003).
[5] G.-H. Kim, A. Pesaran, R. Spotnitz, Journal of Power Sources, 170 (2007) 476-489.
[6] Y.X. Yao, C. Yan, Q. Zhang, Chem Commun (Camb), 56 (2020) 14570-14584.
[7] S. Santhanagopalan, P. Ramadass, J. Zhang, Journal of Power Sources, 194 (2009) 550-557.

[8] L. Terborg, S. Weber, F. Blaske, S. Passerini, M. Winter, U. Karst, S. Nowak, Journal of Power Sources, 242 (2013) 832-837.
[9] D. Liu, J. Pang, J. Zhou, Y. Peng, M. Pecht, Microelectron. Reliab 53 (2013) 832-839.
[10] D. Petz, M.J. Mühlbauer, V. Baran, A. Schökel, V. Kochetov, M. Hofmann, V. Dyadkin, P. Staron, G. Vaughan, U. Lienert, P. Müller-Buschbaum, A. Senyshyn, Energy Storage Materials, 41 (2021) 546-553.
[11] V. Kraft, M. Grutzke, W. Weber, M. Winter, S. Nowak, J Chromatogr A, 1354 (2014) 92-100.
[12] A. Friesen, X. Mönnighoff, M. Börner, J. Haetge, F.M. Schappacher, M. Winter, Journal of Power Sources, 342 (2017) 88-97.
[13] V. Kraft, M. Grutzke, W. Weber, J. Menzel, S. Wiemers-Meyer, M. Winter, S. Nowak, J Chromatogr A, 1409 (2015) 201-209.
[14] M. Grützke, X. Mönnighoff, F. Horsthemke, V. Kraft, M. Winter, S. Nowak, RSC Advances, 5 (2015) 43209-43217.
[15] G. Gachot, S. Grugeon, G.G. Eshetu, D. Mathiron, P. Ribière, M. Armand, S. Laruelle, Electrochimica Acta, 83 (2012) 402-409.
[16] S. Krueger, R. Kloepsch, J. Li, S. Nowak, S. Passerini, M. Winter, Journal of The Electrochemical Society, 160 (2013) A542-A548.
[17] D.R. Gallus, R. Wagner, S. Wiemers-Meyer, M. Winter, I. Cekic-Laskovic, Electrochimica Acta, 184 (2015) 410-416.
[18] J.E. Harlow, S.L. Glazier, J. Li, J.R. Dahn, Journal of The Electrochemical Society, 165 (2018) A3595-A3601.
[19] H. Kato, Y. Kobayashi, H. Miyashiro, Journal of Power Sources, 398 (2018) 49-54.
[20] M.D. Levi, D. Aurbach, Impedance of a single intercalation particle and of non-homogeneous, multilayered porous composite electrodes for Li-ion batteries, J. Phys. Chem. B 108 (2004) 11693–11703.
[21] J.Zhoua, P. H. L. Notten, Journal of The Electrochemical Society, 151 (2004) A2173-A2179.
[22] T. Momma, M. Matsunaga, D. Mukoyama, T. Osaka, J. Power Sources 216 (2012) 304-307.
[23] P. S. Sabet, A. J. Warnecke, F. Meier, Journal of Power Sources, (2019) 227369.
[24] X. Zhou, J. Huang, Z. Pan, M. Ouyang, Journal of Power Sources, 426, (2019) 216-222.
[25] X. Zhou, Z. Pan, X. Han, L. Lu, M. Ouyang, Journal of Power Sources, 417 (2019) 188-192.
[26] Ting Hei Wan, Mattia Saccoccioa, Chi Chena, Francesco Ciuccia, Electrochimica Acta,184 (2015) 483-499
[27] Sara Drvarič, Talian Gregor Kapun, Jože Moškon, Robert Dominko, Miran Gaberšček, Journal of The Electrochemical Society,169 (2022) 010529.
[28] JožeMoškon, Jan Žuntar, Sara Drvarič Talian, Robert Dominko, Miran Gaberšček, Journal of The Electrochemical Society, 167 (2020) 140539
[29] R.DE LEVIE, Elecrrochimica Acta, 8 (1963) 751-780.
[30] R.DE LEVIE, Elecrrochimica Acta, 9 (1964) 1231-1245.
[31] F.Grimsmann, T.Gerbert, F.Brauchle, Gruhlea,J.Parisi, M.Knipper,Journal of Energy Storage, 15 (2018) 17-22.
[32] P. Pfluger, V. Geiser, S. Stolz, H.J. Güntherodt, Synthetic Metals, 3 (1981) 27-39.
[33] C. Hogrefe, T. Waldmann, M. Hölzle, M. Wohlfahrt-Mehrens, Journal of Power Sources, 556 (2023).
[34] K. Smith, Y. Li, F. Piccinini, G. Csucs, C. Balazs, A. Bevilacqua, P. Horvath, Nat Methods, 12 (2015) 404-406.

[35] Wei Yang, Haimei Xie, Baoqin Shi, Haibin Song, Wei Qiu, Qian Zhang, Journal of Power Sources, 423 (2019) 174-182,

[36] J. Illig, M. Ender, T. Chrobak, J. P. Schmidt, D. Klotz, E. Ivers-Tiffee, Journal of The Electrochemical Society, 159 (7) A952-A960 (2012)

[37] Jiangong Zhu, Michael Knapp, Xinyang Liu, Peng Yan, Haifeng Dai, Xuezhe Wei, Helmut Ehrenberg, IEEE Transactions on Transportation Electrification, 7 (2021) 410421.

[38] S.Gantenbein, M. Weiss, E.Ivers-Tiffée, Journal of Power Sources, 379 (2018) 317-327.

[39] R He, Y He, W Xie, B Guo, S Yang, Energy,263(2023) 1703154.

[40] Hahn M, Schindler S, Triebs LC, Danzer MA, Batteries 5 (2019)

[41] Ivers-Tiffee´ E, Weber A, J Ceram Soc Japan, 125 (2017) 193–201.

[42] Shafiei Sabet P, Sauer DU, J Power Sources, 425 (2019)121–9.

[43] W. Li, X.Liu, H. Celio, P. Smith, A.Dolocan, M.Chi, A.Manthiram, Advanced Energy Materials,8 (2018) 1703154.

[44] S.K.Jung, H.Gwon, J.Hong, K.Y.Park, D. H.Seo, H.Kim, J.Hyun, W.Yang, K.Kang, Advanced Energy Materials, 4 (2014) 1300787.

[45] Norihiro Togasaki, Tokihiko Yokoshima, Yasumasa Oguma, Tetsuya Osaka, Journal of Power Sources, 461 (2020) 228168.

[46] Lei Xu, Ye Xiao, Yi Yang, Shi Jie Yang, Xiao Ru Chen, Rui Xu, Yu Xing Yao, Wen Long Cai, Chong Yan, Jia Qi Huang, Qiang Zhang, Angewandte Chemie International Edition 61 (2022) e202210365.

[47] J.P. Schmidt, T. Chrobak, M. Ender, J. Illig, D. Klotz, E. Ivers-Tiffée, Journal of Power Sources, 196 (2011) 5342-5348.

[48] Sheng S. Zhang, Journal of Energy Chemistry 41 (2020) 135–141.

[49] Suting Weng, Siyuan Wu, Zepeng Liu, Carbon Energy 5.1 (2023) e224.

[50] Nan Qin,Liming Jin,Guangguang Xing, Qiang Wu,Junsheng Zheng, Cunman Zhang, Zonghai Chen,Jim P. Zheng， Advanced Energy Materials 13.11 (2023) 2204077.

[51] B. A. Boukamp,Journal of The Electrochemical Society, 142 (1995) 1885.

[52] Rong He, Yongling He, Wenlong Xie, Bin Guo, Shichun Yang, Electrochimica Acta, 444 (2023) 142048.

[53] R. De Levie, Electrochimica Acta, 10 (1965) 113-130.

[54] J. Moškon, M. Gaberšček, Journal of Power Sources Advances, 7 (2021) 100047.

[55] Z. Feng, K. Higa, K. Sung Han, V. Srinivasan, Journal of The Electrochemical Society, 164 (2017) A2434-A2440.

[56] Kazuma Kumai, Hajime Miyashiro, Yo Kobayashi, Katsuhito Takei, Rikio Ishikawa, Journal of Power Sources 81-82 (1999) 715-719.

[57] Xuekun Lu, Marco Lagnoni, Antonio Bertei, Supratim Das, Rhodri E. Owen, Qi Li, Kieran O'Regan, Aaron Wade, Donal P. Finegan, Emma Kendrick, Martin Z. Bazant, Dan J. L. Brett, Paul R. Shearing, Nat Commun 14 (2023) 5127.

[58] D. Petz, M.J. Mühlbauer, A. Schökel, K. Achterhold, F. Pfeiffer, T. Pirling, Batteries & Supercaps, 4 (2021) 327 –335

[59] D. Petz a, M.J. Mühlbauer, V. Baran, M. Frost, A. Sch€okel, C. Paulmann, Y. Chen, Journal of Power